\documentclass{aa}

\usepackage{graphicx}
\usepackage{txfonts}
\usepackage{multirow}
\usepackage[normalem]{ulem}
\usepackage{amstext}
\usepackage{algpseudocode}
\usepackage{hyperref}
\hypersetup{colorlinks = true, citecolor = {blue}}

\usepackage{color}
\usepackage{accents}

\makeatletter
\renewcommand*\aa@pageof{, page \thepage{} of \pageref*{LastPage}}
\makeatother
\newcommand{\msun}{\mathrm{M}_\odot}
\def\PGPU{$\varphi-$GPU }

\def\gapprox{\;\rlap{\lower 3.0pt                       % approximately smaller
        \hbox{$\sim$}}\raise 2.5pt\hbox{$>$}\;}
\def\lapprox{\;\rlap{\lower 3.1pt                       % approximately smaller
        \hbox{$\sim$}}\raise 2.7pt\hbox{$<$}\;}

\newcommand{\be}{ \begin{equation} }
\newcommand{\ee}{\end{equation}}

\newcommand{\ben}{\begin{enumerate}}
\newcommand{\een}{\end{enumerate}}

\newcommand{\orcid}[1]{\href{https://orcid.org/#1}{\protect\includegraphics[width=8pt]{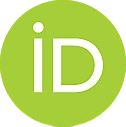}}}

\begin{document}

\title{Star-by-star dynamical evolution of the physical pair of the Collinder~135 and UBC~7 open clusters}

\titlerunning{Cr~135 and UBC 7: Simulations}

\author{
Maryna~Ishchenko
\inst{1,2}\fnmsep\thanks{\tt marina@mao.kiev.ua}
\orcid{0000-0002-6961-8170}
\and
Dana~A.~Kovaleva
\inst{} 
\orcid{0000-0003-4328-8801}
\and
Peter~Berczik
\inst{1,2,3,5}
\orcid{0000-0003-4176-152X}
\and
Nina~V.~Kharchenko
\inst{}
\orcid{0000-0003-4584-6651}
\and
Anatoly E. Piskunov
\inst{}
\orcid{0000-0003-4110-7915}
\and
Evgeny Polyachenko
\inst{3,1}
\orcid{0000-0002-4596-1222}
\and
Ekaterina~Postnikova
\inst{}
\orcid{0009-0009-3481-441X}
\and
Andreas~Just
\inst{3}
\orcid{0000-0002-5144-9233}
\and
Olga~Borodina
\inst{4}
\orcid{0000-0001-7522-1324}
\and
Chingis~Omarov
\inst{2}
\orcid{0000-0002-1672-894X}
\and
Olexandr~Sobodar
\inst{1}
\orcid{0000-0001-5788-9996}
}

\institute{Nicolaus Copernicus Astronomical Centre, Polish Academy of Sciences, ul. Bartycka 18, 00-716 Warsaw, Poland
           \and 
Fesenkov Astrophysical Institute, 050020, Almaty, Kazakhstan
           \and
%Main Astronomical Observatory, National %Academy of Sciences of Ukraine, 27 %Akademika Zabolotnoho St, 03143 Kyiv, %Ukraine             
%           \and
Zentrum f\"ur Astronomie der Universit\"at Heidelberg, Astronomisches Rechen-Institut, M\"onchhofstr 12-14, 69120 Heidelberg, Germany
           \and
Center for Astrophysics $\vert$ Harvard \& Smithsonian, 60 Garden St, Cambridge, MA 02138, USA
        \and
Konkoly Observatory, HUN-REN Research Centre for Astronomy and Earth Sciences, Konkoly Thege Mikl\'os \'ut 15-17, 1121 Budapest, Hungary
          }
   
\date{Received ... / Accepted ...}

\abstract    
% context heading (optional)
{In a previous paper using \textit{Gaia} DR2 data, we demonstrated that the two closely situated open clusters Collinder 135 and UBC 7 might have formed together about 50 Myr ago.}
% aims heading (mandatory)
{In this work, we performed star-by-star dynamical modelling of the evolution of the open clusters Collinder 135 and UBC 7 from their supposed initial state to their present-day state, reproducing observational distributions of members.}
% methods heading (mandatory)
{Modelling of the Collinder 135 and UBC 7 dynamical evolution was done using the high-order parallel $N$-body code \PGPU with up-to-date stellar evolution. Membership and characteristics of the clusters were acquired based on Gaia DR3 data.}
% results heading (mandatory)
{The comparison of the present-day radial cumulative star count obtained from the $N$-body simulations with the current observational data gave us full consistency of the model with observational data, especially in the central 8 pc, where 80\% of the stars reside. The proper motion velocity components obtained from the $N$-body simulations of the stars are also quite consistent with the observed distributions and error bars.}
% conclusions heading (optional), leave it empty if necessary
{These results show that our numerical modelling is able to reproduce the open clusters' current complex 6D observed phase-space distributions with a high level of confidence. Thus, the model demonstrates that the hypothesis of a common origin of Collinder~135 and UBC~7 complies with present-day observational data.}

     
\maketitle

%%%%%%%%%%%%%%%%%%%%%%%%%%%%%%%%%%%%%%%%%%%%%%%%%%%%%%%%%%%%%%%%%%%%
\section{Introduction}% 
\label{sec:Intr}
%%%%%%%%%%%%%%%%%%%%%%%%%%%%%%%%%%%%%%%%%%%%%%%%%%%%%%%%%%%%%%%%%%%%
The data of the space mission Gaia \citep{2016A&A...595A...1G} have stimulated investigations into open clusters (hereafter OCs) in the Galaxy. One of the areas of research that has significantly benefited is related to the binarity of OCs. While a lack of observational data of binary star clusters in our Galaxy in comparison with the Magellanic Clouds has been discussed previously \citep[e.g.][]{2010A&A...511A..38V, 2009A&A...500L..13D}, 
recent investigations have provided a number of examples of pairs of OCs in our Galaxy that suggest a common origin \citep[see, e.g.][]{2023ApJS..265...12Q, 2022MNRAS.510.5695A, 2022AJ....164..132Y, 2022Univ....8..368C, 2021MNRAS.503.5929B, 2019A&A...624A..34Z}.
Some clusters that were previously considered single have shown a binary signature upon detailed examination \citep []{2020A&A...633A.146A, 2018MNRAS.481.3887D}, including clusters with signs of merging by tidal capture \citep []{2021ApJ...923...21C}, according to Gaia data.

There is observational evidence that indicates binary star clusters are usually coeval and form as a result of hierarchical clustering in parental molecular clouds \citep[]{2022Univ....8..113C, 2022MNRAS.510.5695A}. These pairs may or may not be bound and physically interacting. Notably, \cite{2022MNRAS.510.5695A} found four bound and two interacting but unbound  pairs of star clusters. Alternatively, coeval clusters may be located relatively close to each other due to being born in the same star forming complex but may never interact \citep{2022Univ....8..113C, 2022Univ....8..368C, 2023MNRAS.521.1399C}, and eventually they may become disaggregated while maintaining similar Galactic orbits \citep{2022ApJ...928...70C}.  On the other hand, pairs of closely located OCs with significantly different ages, which represent bona fide events of an occasional encounter, have been found \citep{2021ApJ...923...21C, 2022MNRAS.511L...1P, 2023A&A...675A..28O}. Dynamical simulation of such encounters has so far been limited to the publicly available orbit integration code {\it galpy} \citep{2015ApJS..216...29B}. The dynamical simulations of the evolution of initially bound primordial OCs in the Galactic tidal field by \cite{2010ApJ...719..104D} have demonstrated that very close pairs, if formed, are merged in a short timescale, and in other pairs, either the smaller component is disrupted or a pair is decoupled in the background tidal field due to mass loss, making open star cluster binarity a transient phenomenon. \cite{2016MNRAS.457.1339P} used $N$-body simulations to explore the evolution of binary open star clusters in the Galactic tidal field with different initial orbital orientations with respect to the Galactic plane, which proved to be another important factor influencing the dynamical fate of such clusters. \cite{2021MNRAS.506.4603D} carried out $N$-body simulations to investigate the formation and evolution of binary star clusters in the Milky Way and in the Large Magellanic Cloud and found that binary open star clusters can form from stellar aggregates with a variety of initial conditions.

In \cite{2020A&A...642L...4K}, hereafter \hyperlink{K20}{\color{blue}{Paper~I}}, we discussed possible scenarios of origin and dynamical evolution in a simplified approach and constrained parameters, within the limits of observational values, that suggest a common origin of the open star clusters Collinder 135 (hereafter Cr~135) and UBC 7 \citep[see also discussion in][]{2021ApJ...923...20P}. As one of the results of that work, we identified, using backward orbital integration, a presumable grouping of initial states in 6D space that corresponds to a common origin of these clusters. Our previous work was limited only by the investigation of motion of OC dynamical centres (centre of masses). So, we completely neglected the OC internal dynamical and stellar evolution, including the mass loss of the clusters. By overcoming such restrictions in our current investigation, we make our previous results more robust and reliable. 

In this article, our purpose is to reproduce the dynamical evolution of the system of Cr~135 and UBC~7 using star-by-star simulations. During our investigation, we especially paid attention to the complex dynamical co-evolution and cross-penetration of the clusters' stellar populations. This behaviour of the modelled stellar systems made our attempts of system modelling quite challenging. 

In comparison with  \hyperlink{K20}{\color{blue}{Paper~I}}, we have updated the numerical implementation of the \PGPU with a new stellar evolution block \citep{Banerjee2020, Kamlah2022} combined with our high-order dynamical $N$-body code. This new implementation allowed us to  predict with a good degree of confidence the evolved clusters' internal stellar structure and cumulative mass distribution.

After the publication of \hyperlink{K20}{\color{blue}{Paper~I}}, the \textit{Gaia} DR3 catalogue \citep{2021A&A...649A...1G, 2023A&A...674A...1G} was published. With respect to the previously used \textit{Gaia} DR2 catalogue \citep{2018A&A...616A...1G}, \textit{Gaia} DR3 provides enhanced astrometric and photometric data for an enlarged sample of the stars as well as a significantly larger number of sources with mean radial velocities. Some systematic effects have been resolved or mitigated regarding \textit{Gaia} DR2, which is also beneficial for the study of the population of OCs. These improvements allowed us to obtain refined characteristics of the system composed of the two clusters based on the deep census of the probable members of this system.  

Our goal is to reproduce the present-day state of OCs and compare it with the most up-to-date information about their stellar population from \textit{Gaia} DR3 data. Such an ambitious investigation has only been possible since last few years thanks to the Gaia's unprecedented position and proper motion measurements. As a possible physical pair of OCs, Cr~135 and UBC~7 give us a unique laboratory to also study the common processes of the ongoing star formation on larger scales in our Milky Way disc. 

The paper is organised as follows. In Sec.~\ref{sec:dyn}, we describe dynamical $N$-body modelling of evolution of the pair of clusters. In Section~\ref{sec:memb}, we discuss the data filtering process that resulted in the following view onto the system of the two OCs, Cr~135 and UBC~7, with {\it Gaia} DR3 data. Section~\ref{sec:sys} is focused on characterisation of the system based on the obtained characteristics of its members. In Section~\ref{sec:disc}, we describe results obtained in our $N$-body simulations in comparison with the observational data presented in Sections~\ref{sec:memb} and \ref{sec:sys}. In Section \ref{sec:disc}, we also discuss kinematics of the surrounding extended halo. In Section~\ref{sec:conc}, we sum up the main results of the work.

%%%%%%%%%%%%%%%%%%%%%%%%%%%%%%%%%%%%%%%%%%%%%%%%%%%%%%%%%%%%%%%%%%%%%

\section{Cr 135 and UBC 7: Star-by-star N-body modelling}
\label{sec:dyn}

\subsection{Initial conditions and integration procedure}

In \hyperlink{K20}{\color{blue}{Paper~I}} we found that it is highly probable that Cr 135 and UBC 7 form a physical pair. This conclusion is based on backward-in-time numerical simulations and 6D observations from the \textit{Gaia} DR2 catalogue. {The resulting initial positions and velocities for cluster centres of mass (initial separation and velocity: 10 pc and 0.95 km/s) showed that the probability of a mutual formation scenario is about 98\% (see the numerical model {\tt \#[53,61]}, and Fig. 2 of \hyperlink{K20}{\color{blue}{Paper~I}})}. In order to estimate the impact of {\it Gaia} DR3 data on new simulations, we used newly determined present-day cluster observed parameters determined as described in detail in Sec.~\ref{sec:memb}. We found that there is good agreement in the newly determined Galactic coordinates, proper motions, and radial velocities (see Table~\ref{tab:data_dr2_3}) with the {\it Gaia} DR2 parameters derived in \hyperlink{K20}{\color{blue}{Paper~I}}). 

\begin{table}[tbp]
\setlength{\tabcolsep}{2pt}
%\centering
\caption{Comparison of the main parameters of the OCs based on two \textit{Gaia} data releases.}
\label{tab:data_dr2_3}
\begin{tabular}{cccccccc}
\hline 
\hline 
%DR & SC &  $l$, deg & $b$, deg & $\varpi$, mas &$\mu_l^*$, mas/yr & $\mu_b$, mas/yr & $V_r$, km/s \\
DR & OC &  $l$, & $b$, & $\varpi$, &$\mu_l^*$,& $\mu_b$,& $V_r$,\\
 & &  deg & deg & mas &mas/yr & mas/yr &  km/s \\
\hline
3 & Cr~135 & 248.75 & $-$11.10 &3.38 &$-$9.99  & $-$6.40 & 15.2  \\
 & &$\pm$0.06& $\pm$0.04 &$\pm$0.03 &$\pm$0.03 & $\pm$0.03  & $\pm$1.5 \\
2 & Cr~135 & 248.98 & $-$11.10 &3.31  &$-$9.92  & $-$6.47 & 17.4  \\
 &   & $\pm$0.06& $\pm$0.05& $\pm$ 0.02&$\pm$ 0.05 &$\pm$  0.06& $\pm$ 1.3 \\
%* & Cr~135 & +248.9760 & -11.0982 & -9.9180 & -6.4680 & 17.4500 \\

3 & UBC~7  & 248.62 & $-$13.45 &3.61 &$-$10.38 & $-$5.90 & 15.8 \\
 &   & $\pm$0.04& $\pm$0.05& $\pm$0.03 &$\pm$ 0.03 &$\pm$ 0.03 & $\pm$ 2.8 \\
2 & UBC~7  & 248.62 & $-$13.37 & 3.56&$-$10.25 & $-$5.98 & 16.7 \\
 &   & $\pm$0.04 & $\pm$ 0.05& $\pm$ 0.02   & $\pm$0.05  & $\pm$ 0.05 & $\pm$ 1.5 \\
%* & UBC~7  & +248.6240 & -13.3700 & -10.2470 & -5.9770 & 16.7100 \\
\hline 
\end{tabular} 
\vspace{1pt}
\\ $^*$ -- here and further $\mu_l \equiv \mu_l^0\cdot \cos b$
\end{table}

\begin{figure}
\includegraphics[width=0.99\linewidth]{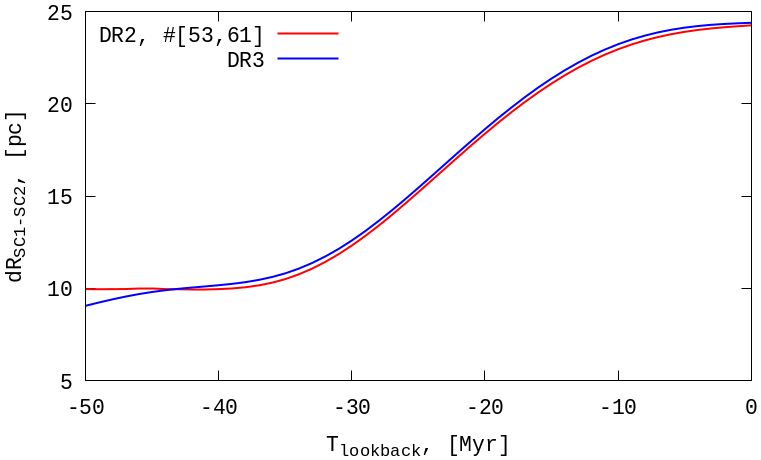}
\caption{Evolution of the distance between the centres of Cr~135 and UBC~7 for the DR2 (red) and DR3 (blue) initial conditions.}
\label{fig:dccorr-dr2-3}
\end{figure}

{To evaluate the possible distinction (DR2 vs. DR3) in the  distances between the two OC centres, we carried out a new lookback integration using the new DR3 initial data from Table~\ref{tab:data_dr2_3} and compared it to our earlier selected best numerical model {\tt \#[53,61]} (see Fig. \ref{fig:dccorr-dr2-3}).} As can be seen in the figure, the agreement is good, and the effect of the different data releases is small.

For the OCs' dynamical orbital integration including the stellar evolution, we used the high-order parallel $N$-body code \PGPU,\footnote{$N$-body code \PGPU: \\~\url{ https://github.com/berczik/phi-GPU-mole}.} which is based on the fourth-order Hermite integration scheme with individual hierarchical block time steps \citep{Berczik2011,BSW2013}. The current version of the \PGPU code uses a native GPU support and a direct code access to the GPU using the NVIDIA native CUDA library. The present code is well tested and was used to obtain important results in our earlier OC simulations \cite{Just2009, Shukirgaliyev2017, Shukirgaliyev2018, Shukirgaliyev2021}. 

Because of the relatively short evolutionary timescale of our system, we used the so-called fixed external potential with parameters independent of time. Similar to \cite{Kharchenko2009, Ishchenko2023}, we used the three-component (bulge-disc-halo) axisymmetric Plummer-Kuzmin model \citep{Miyamoto1975}.

The current code also takes into account up-to-date stellar evolution models. The most important updates were made to components of stellar evolution and include updated metallicity dependent stellar winds; updated metallicity dependent core-collapse supernovae, along with a new fallback prescription and their remnant masses; updated electron-capture supernovae accretion induced collapse and merger-induced collapse; remnant masses and natal kicks; and black hole natal spins, among other updates. (For more details about the new stellar evolution library, see \cite{Kamlah2022}).

To carry out the star-by-star $N$-body modelling together with the stellar evolution for both OCs, we assumed the clusters to be in initial dynamical equilibrium. For the star particles, we used the King model distribution function \cite{King_1966} with three free parameters: the initial total mass, the initial half-mass radius $r_{\rm hm}$, and the dimensionless central potential $W_0$. Using the Kroupa \citep{2001MNRAS.322..231K} initial mass function with the lower and upper mass limits equal to 0.1 -- 10 $\msun$, and the classical solar value for the initial stellar metallicity, $Z = Z_\odot = 0.02$ \citep{Grevesse1998}, for both clusters. For the centres of the clusters, we used the initial coordinates and velocities at the 50 Myr lookback time integration (see Table \ref{tab:data1}). 

We tested more than 50 runs with various combinations of $M, r_{\rm hm}$, and $W_0$ separately for each cluster using the full $N$-body model calculations up to the present time. For the initial positions and orbital velocities in Cartesian Galactic coordinates of the star clusters (SC) centre at the -50 Myr lookback in time, we used the values obtained during our own set of lookback integration of the SC centre of mass. The initial masses, radii, and King profiles were found during the fitting test runs, with the aim of matching the theoretical and observed parameters. The basic initial parameters of the model are given in Table~\ref{tab:data1} separately for each SC. 

\begin{table*}[tbp]
\setlength{\tabcolsep}{4pt}
\centering
\caption{Initial positions, velocities, and parameters for OCs at 50 Myr lookback time.}
\label{tab:data1}
\begin{tabular}{ccccccccccc}
\hline
\hline 
SC & $X$, pc & $Y$, pc & $Z$, pc & $V_x$, km/s & $V_y$, km/s & $V_z$, km/s & $M$, $\msun$ & $N$ & $r_{\rm hm}$, pc & $W_0$ \\
\hline
Cr~135 & --1073.5 & --8415.3 & --20.4 & --230.17 & 34.14 & --5.38 & 230 & 442 & 10 & 3 \\
UBC~7  & --1076.0 & --8417.7 & --12.0 & --230.31 & 34.96 & --5.76 & 200 & 384 & 7 & 11\\
\hline 
\end{tabular}
\vspace{6pt}
\end{table*}

\begin{figure}
\includegraphics[width=0.99\linewidth]{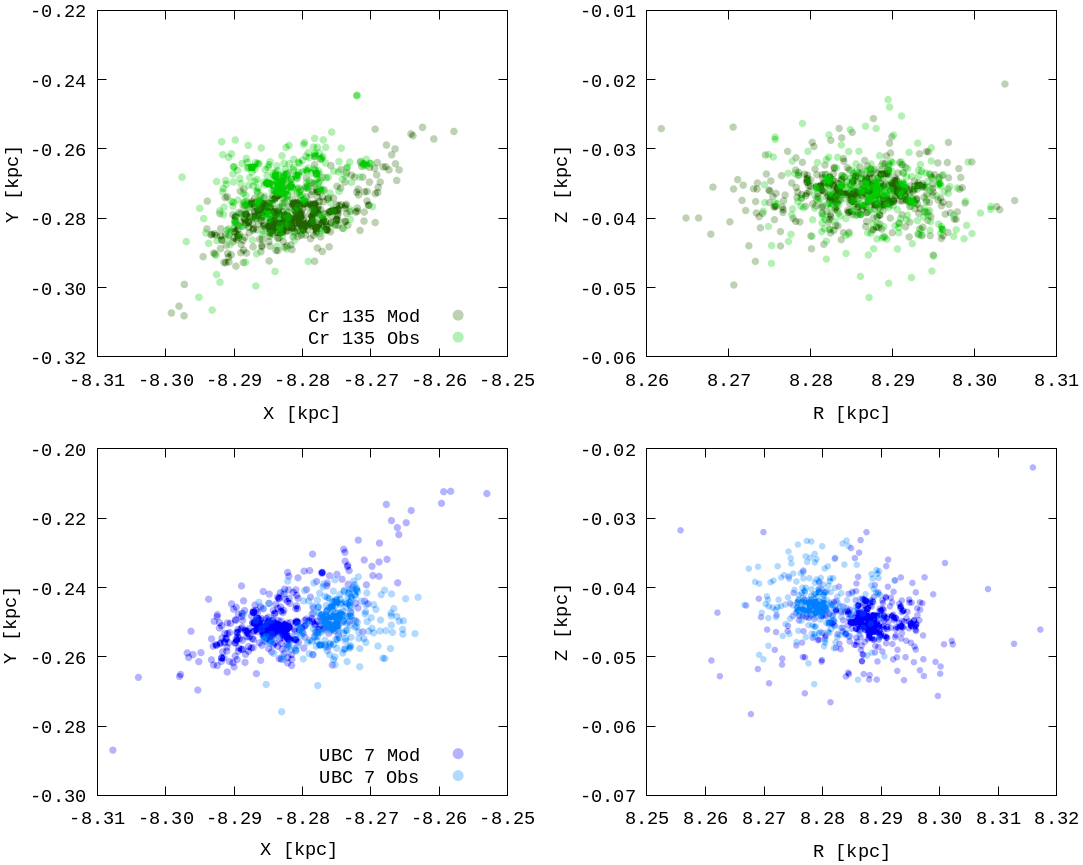}
\caption{Comparison of observed positions of probable cluster members and of model stars of basic $N$-body simulations. The light green and light blue dots are the members of Cr~135 and UBC~7, respectively. The coordinate frame was obtained by parallel translation of the standard Galactic coordinate system.} 
\label{fig:sc-poss}
\end{figure}

Figure~\ref{fig:sc-poss} shows the observed stars of the OCs (light green for Cr 135 and light blue for UBC 7) overplotted with the present-day positions of all particles from our basic model according to the initial data in Table \ref{tab:data1} (dark green for Cr 135 and dark blue for UBC 7). The mutual orientation of the OCs is correct, although both sets of particles (representing Cr~135 and UBC~7) are slightly misaligned with their prototype clusters (see Table \ref{tab:data2}). In Table \ref{tab:data2}, we present the positions and velocities for the current centres of mass for both OCs. The identifier {\tt Sim} indicates the cluster's centre of mass as obtained from our basic model, and {\tt Obs} indicates the centre of mass as obtained from observations (see Section~\ref{sec:memb}). The small disagreement  between the 3D coordinates and velocities inside the table most probably arises from the different mutual interactions of the clusters in the backward integration of the orbit compared to the forward integration of the full OCs. The backward integration was performed with the present-day masses of the OCs. The forward integration started with the larger initial cluster masses and includes mass loss by stellar evolution and the tidal field. The present-day shift of the cluster centres demonstrates that Cr 135 and UBC 7 interact gravitationally with each other.

\begin{table}[tbp]
\setlength{\tabcolsep}{1pt}
\centering
\caption{Comparison of the observed and numerical centre-of-mass positions and velocities for Cr 135 and UBC 7 in Cartesian Galactic coordinates for the present day.}
\label{tab:data2}
\begin{tabular}{ccccccccccc}
\hline 
\hline 
SC  &  & $X$, pc & $Y$, pc & $Z$, pc & $V_x$, km/s & $V_y$, km/s & $V_z$, km/s \\
\hline
Cr~135 & Sim & --8281.3 & --277.0 & --36.2 & --7.63 & 237.35 & --5.11 \\
       & Obs & --8282.1 & --271.0 & --36.2 & --7.38 & 237.94 & --4.85\\
UBC~7 & Sim & --8283.1 & --239.5 & --45.2 &     --7.10 & 237.71 & --4.72 \\
      & Obs & --8276.2 & --250.8 & --43.6 & --7.00 & 238.13 & --4.27\\
\hline 
\end{tabular}
\vspace{6pt}
\end{table}

\subsection{Stellar evolution and lifetime}

\begin{figure}
\includegraphics[width=0.99\linewidth]{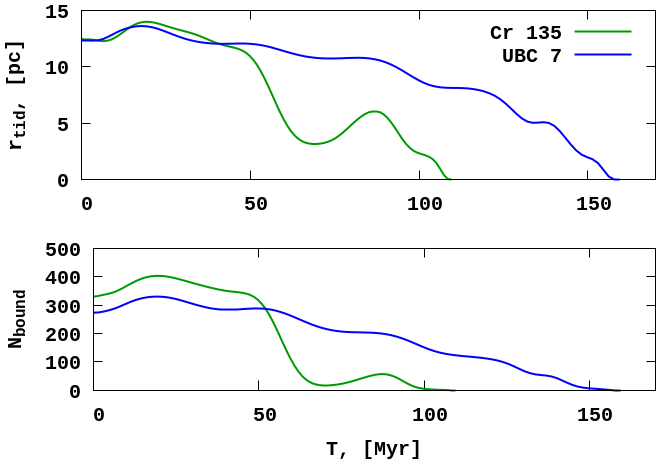}
\caption{Evolution of the tidal radius (\textit{upper panel}) and bound particle number (\textit{bottom panel}). The green line represents Cr 135, and the blue line indicates UBC 7.} 
\label{fig:tid}
\end{figure}

In Fig. \ref{fig:tid}, we present the evolution of the tidal radius and bound particle numbers according to our basic $N$-body model. The radii of both clusters show a slight increase during the first several tens of million years, after which the radius of Cr 135 starts to sharply decrease, and after approximately 110 million years of evolution, the core falls apart {(technically, the number of bound particles decreases to zero)}. At this time, the separation between the two OCs is about 50 pc. UBC~7 has a more gradual decrease in tidal radius in contrast to Cr~135, and it falls apart approximately 50 million years after the dissolution of Cr 135. The differences between the dynamical behaviour of our clusters can be explained as being a result of the concentration parameters $W_0$ = 3 and $W_0$ = 11 of the different initial King profiles. Thus, for Cr 135 the  profile indicates a looser and more unstable core. In UBC~7, the core is more concentrated, and therefore, despite the smaller number of stars, it demonstrates a greater ability to survive. 

\begin{figure}
\includegraphics[width=0.99\linewidth]{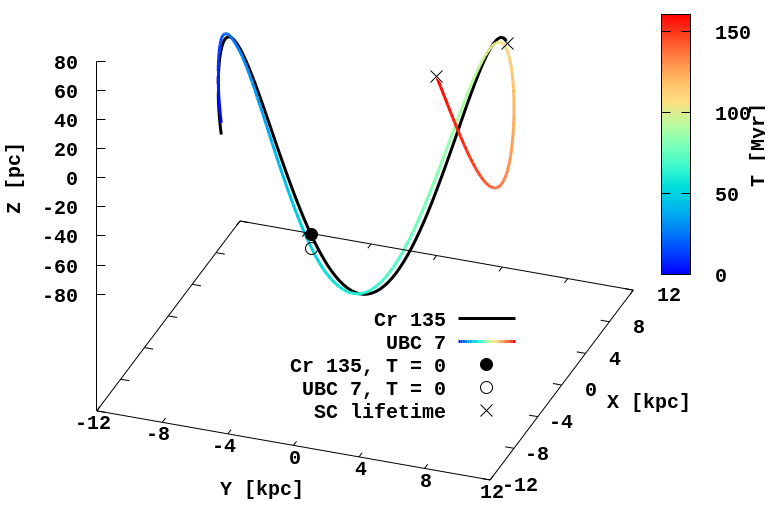}
\caption{Orbital evolution of Cr~135 and UBC~7. The colour-coding represents the time of integration. The filled and open black circles indicate the present positions of the clusters. Th black crosses indicate the positions where the clusters dissolve.} 
\label{fig:orb-evol}
\end{figure}

In Fig. \ref{fig:orb-evol}, we show the orbital evolution up to $\sim$150 Myr from the formation of the OCs. As can be seen from Fig. \ref{fig:orb-evol}, during the time of evolution, the clusters follow mildly eccentric clockwise orbits around the Galactic centre, as seen from the Galactic north pole, and simultaneously rotate around each other. The orbits have considerable vertical oscillations in the $z$-direction between $\pm 80$ pc. 

\begin{figure}
 \includegraphics[width=0.99\linewidth]{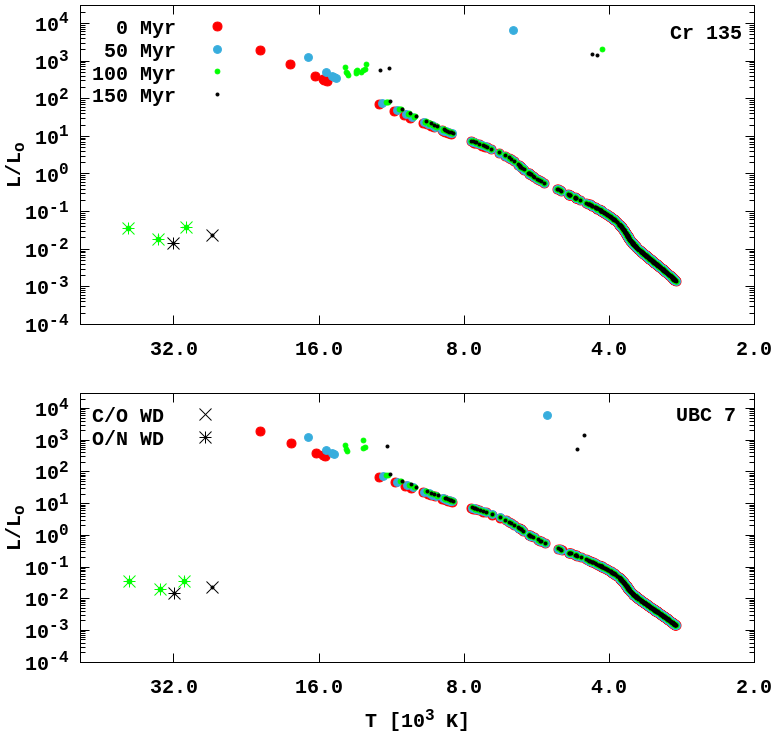}
\caption{Hertzsprung-Russell diagrams of Cr~135 and UBC~7 (including all unbound stars). Red dots correspond to an age of 0 Myr, light blue to 50 Myr, green to 100 Myr, and black to 150 Myr. The crosses and the stars show two types of white dwarfs from the $N$-body simulation.}  
\label{fig:tl-diag}
\end{figure}

In Fig. \ref{fig:tl-diag}, we present the Hertzsprung-Russell diagram (HRD) for three moments of evolution (50, 100, and 150 Myr in forward integration) for each OC starting from 0 Myr. Due to the low number of available stars and the short period of the OCs' lifetime, we mainly have main-sequence stars. When analysing the type of stars for the present day (which have an age of 50 Myr according to the $N$-body simulation), $\sim 84\%$ of them are low-mass stars (m < 0.8 M$_\odot$), and only $\sim16\%$ of them are higher-mass main-sequence stars for both OCs. At the end of the forward integration, low-mass stars were still dominant, and there were only a few carbon/oxygen and oxygen/neon white dwarf stars.

%=================================================================================
\section{Cluster membership with Gaia DR3 data}
\label{sec:memb}
%=================================================================================

Our work in \hyperlink{K20}{\color{blue}{Paper~I}} is based on {\it Gaia} DR2 data and aimed at selecting the most probable members of Cr~135 and UBC~7 in order to obtain their global space and kinematic parameters. We incidentally discovered an extended corona of stars with an almost constant projected density inside a 20-pc radius around the centre of mass of both clusters. In the present investigation, our aim is to use the wealth of the {\it Gaia} DR3 data to explore stellar membership and the structure of the whole system in detail.

Recent investigations have proved the existence of extended features connected with groups of OC (see, for example, \cite{2020MNRAS.491.2205B, 2019A&A...624A..34Z, 2020ApJ...900L...4P}). On the other hand, the discussed Cr~135 and UBC~7 are situated in the Vela-Puppis star-formation region (Galactic longitude $245\degr \lesssim l \lesssim 265\degr$ and latitude 
$-15^\circ \lesssim b \lesssim -5^\circ$), which hosts several young OCs with ages between approximately 15 and 50 Myr \citep{2019A&A...621A.115C, 2019A&A...626A..17C, 2020MNRAS.491.2205B}. Of these clusters, Cr~140, NGC~2451B, and NGC~2547 are situated at about 100~pc from the investigated pair. The nearest neighbour to Cr~135 and UBC~7 is the evolved open cluster Alessi 3 \citep[625 Myr,] [] {2023A&A...673A.114H}. 
Thus, in order to reveal the size and characteristics of the coronae and check whether it is due to contamination from neighbouring stellar groups, we considered a very wide region surrounding Cr~135 and UBC~7. This required refining of the data selection principles used in \hyperlink{K20}{\color{blue}{Paper~I}}. 

We performed an intensive search for members related to OCs Cr~135 and UBC~7 among all \textit{Gaia} DR3 sources satisfying the following conditions:
\begin{description}
    \centering
    \item $2.5<\varpi\ {\rm [mas]} <5$, 

\item $220\degr<l<280\degr$, 

\item $-25\degr<b<1\degr$, 

\item $1.2<\mu_{\rm \alpha}^*\ {\rm [mas/yr]}<11.8$, and

\item $-15<\mu_{\rm \delta}\ {\rm [mas/yr]}<-5$. 
\end{description}

The linear size of a region was determined based on previous research in order to include regions that are most probably free of members of central clusters. These external regions are useful for the estimation of random background contamination from the procedure. We applied parallax zero-point correction \citep{2021A&A...649A...4L} with the algorithm provided by \textit{Gaia.}\footnote{\label{note1} \textit{Gaia} EDR3 source code from \\~\url{https://www.cosmos.esa.int/web/gaia/edr3-code}.}

\begin{figure*}
  \includegraphics[width=0.49 \linewidth]{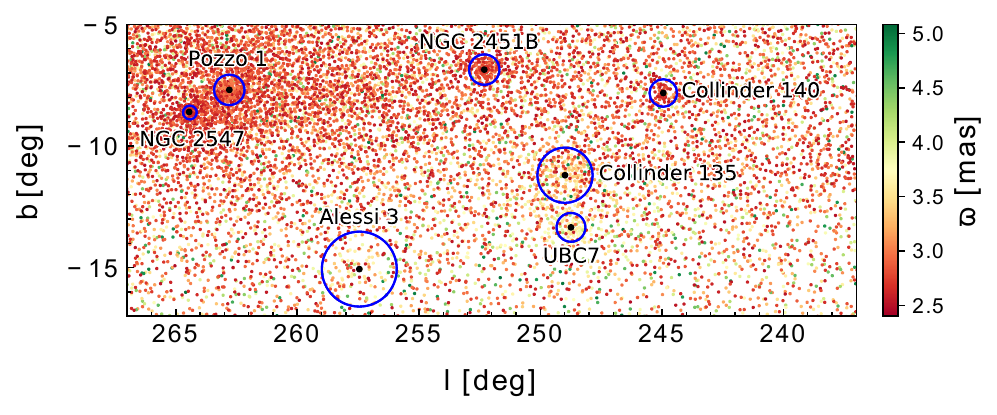}
  \includegraphics[width=0.49 \linewidth]{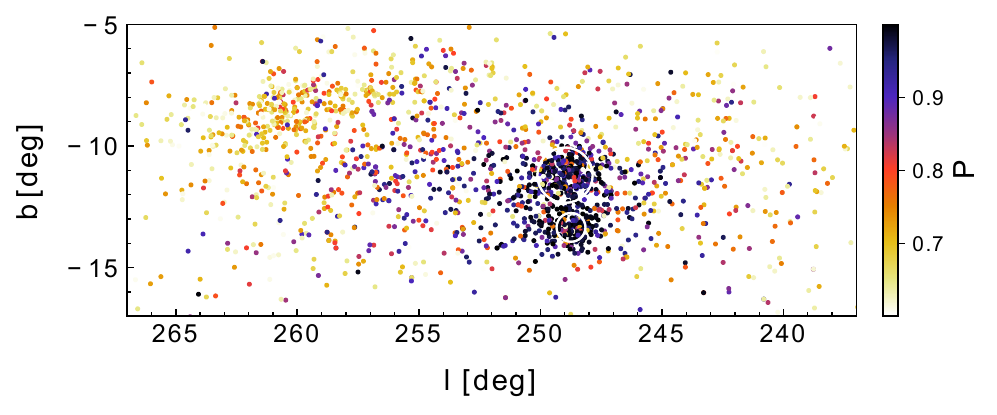}
  \caption{A selection of Gaia DR3 data in the broad vicinity of the Cr~135 and UBC~7. Cr~135, UBC~7, and their closest neighbouring clusters according to \cite{2020A&A...640A...1C}, \textit{left panel}. Refined sample of $60\%$ of the members of Cr~135 and UBC~7 over the same sky area, \textit{right panel}. The colour-coding indicates the MP according to Eq.~\ref{eq:ptot}. The circles in both panels represent the half-number radii of the clusters according to \cite{2020A&A...640A...1C}. See discussion in the text.}
  \label{fig:widestrip}
\end{figure*}

Figure~\ref{fig:widestrip} presents the primary selection containing 116 975 sources. Their colour-coded parallax ranges from yellow, for the most distant stars of the sample at about 400 pc from the observer, to dark violet, for foreground stars at 250 pc from the Sun. Even with this rough selection, some of the OCs of a region became noticeable, Cr~135 and UBC~7 being the nearest two. At a distance of about 280 to 300 pc from the Sun where they are situated, 1~deg $\approx$ 5~pc, so  the linear size of the rectangle in the frontal plane  is about 300 pc in $l$ and about 75 pc in $b$ (Fig.\ref{fig:widestrip}, \textit{left panel}). One can see an enhanced density of sources towards the Galactic plane in $b$ and towards the Vela Molecular Ridge and the Galactic centre in $l$. As we applied no filter to the quality of astrometric and photometric solutions, part of the faint stars may belong to the background and may reflect the stellar density of distant regions \citep{2018A&A...616A...2L}.

We kept as a basis the procedure of probabilistic selection of cluster members described by \cite{2012A&A...543A.156K}. Each star of the considered dataset was assigned a probability $P$ of sharing age, space, and kinematics of a pair of clusters, namely Cr~135 and UBC~7:
\begin{equation}
\label{eq:ptot}
    P=\min(P_{\rm kin}, P_{\rm \varpi}, P_{\rm ph}).
\end{equation}
where $P_{\rm kin}, P_{\rm \varpi}, P_{\rm ph}$ are the probabilities based on kinematics, parallax, and photometric data. The procedure of calculating the probabilities is described in detail in Appendix~\ref{memb}.

After each source in our dataset was assigned a membership probability (MP) according to Eq.~(\ref{eq:ptot}), we selected stars with $P>0.6$ (so-called $60\%$-members) for further consideration. In Fig. \ref{fig:widestrip} (right panel), one can observe two dense central groups (i.e. the cores of Cr~135 and UBC~7) and an extended group of stars with a high MP resembling a halo surrounding the cores. Also present are sparse, chaotically distributed sources of lower probability that dominate the periphery, which presumably are contaminating foreground and background stars, and an elongated dense structure between $250\degr<l<263\degr$ and $-11\degr<b<-7\degr$ formed by members with an intermediate MP. After investigation, we found that this elongated stellar group has a systematically different space motion and does not belong to the system we are focused on. This group may in part be associated with a group of stars identified as a loosely spread cluster [BBJ2018]~6 in \citet{2018MNRAS.481L..11B}, but due to its spatial scale, it might be referred to as a stream instead. For the purposes of the present work, we identified and removed from consideration sources with MPs to the [BBJ2018]~6 group larger than MPs to the group of Cr~135 and UBC~7. The remaining dataset represented in Fig.~\ref{fig:clu_halo_back} consists of 1714 stars. One may observe that the probable members of the group composed of the two clusters still form a distinct asymmetrical halo around the location of the dense clusters.

Random contamination can be estimated based on the number of probable members in the outer parts of the field, which are approximately at $l>270\degr$, $l<235\degr$, $b>-4\degr$, $b<-20\degr$. For this purpose, as well as to set a distinction between the halo and the background, we relied on the density distribution of the stars in the tangential plane. For each source, we estimated its weight over the background $W_i$ as the residual: 

\begin{equation}
    W_i=\rho(l_i,b_i)-\rho_{\rm r}(l_i,b_i),
\end{equation}
\noindent where $\rho(l,b)$ is the stellar density distribution of the refined dataset convolved by the 2D Gaussian kernel of the width of 2\degr and $\rho_{\rm r}(l,b)$ is the similarly convolved random flat distribution of the same number of stars over the same field. 
%For each source we estimate its weight over background $W_i$ as the residual: 
%\begin{equation}
%    W_i=\rho(l_i,b_i)-\rho_{\rm r}(l_i,b_i)
%\end{equation}
We considered the 293 sources with $W_i \leq 0$ to be background contamination stars, and we treated the 1421 stars with a positive $W_i$ as those composing the clusters and the halo. From hereon, we refer to this dataset as Sample I (available via CDS).

Our purpose was to recover the most probable distribution of sources in Sample I between three substructures: Cr~135, UBC~7, and their extended halo. In space, these groups appear to be separated, but there is no distinct border between them. Given that the separation between the centres of the clusters is just 22~pc, peripheral stars of the clusters are mixed and may only be conditionally attributed to one of them. Due to the similarity of their mean kinematic characteristics, we attributed each source to one of these components based solely on its maximum likelihood location in space with respect to the clusters as follows. 

Using the mean position of the reliable members from Table~\ref{tab:data_dr2_3}, we suggested for the clusters a Gaussian 3D PDF $P^{sc}_k$ over a Cartesian distance to the centre with coordinates based on Gaia DR3 ($l^0_k, b^0_k, \varpi^0_k$) from Table~\ref{tab:data_dr2_3}, where $k=1$ refers to Cr~135 and $k=2$ is for UBC~7:

\begin{equation}
\Pi(d_k) = \exp\left[\frac{\left(d^i_k\right)^2}{2\sigma_{d^k}^2}\right]\,,
\label{PDFcluster}
\end{equation}
\noindent 

where $\sigma_{d^k}$ is based on the distribution of the reliable cluster members by Gaia DR3 (Table~\ref{tab:data_dr2_3}). We considered the distance $d$ between the centre of an OC ($l^0_k, b^0_k, \varpi^0_k$) and the $i^{th}$ source of Sample I ($l^i, b^i, \varpi^i$) to be a function of $\varpi^i_m$ with a Gaussian PDF:

\begin{equation}
\Pi(\varpi^i) = \exp\left[-\frac{\left(\varpi^i-\varpi^i_m\right)^2}{2\sigma^2_{\varpi^i}}\right]\,,
\label{PDFparallax}
\end{equation}
\noindent 
where the standard deviation $\sigma^i_\varpi \equiv \epsilon_\varpi^{i}$. 
We simulated numerically the resulting PDF for the distance of each source of Sample I to the cluster centre $\Pi(d_k,\varpi^i) = \Pi(d_k)\cdot \Pi(\varpi^i_k)$, and we compared the  maximum values of PDFs for $k=1,2$:
%\begin{description}
%\centering
\begin{algorithmic}
\If{$\max(\Pi(d_1,\varpi^i)) \geq \max(\Pi(d_2,\varpi^i)) \land d_1^i[\max(\Pi(d_1,\varpi^i))] \leq 15$~pc,} 
\State{the source $i \in$ Cr~135},
%\EndIf
\Else
\If{$\max(\Pi(d_2,\varpi^i)) \geq \max(\Pi(d_1,\varpi^i)) \land d_2^i(\max(\Pi(d_2,\varpi^i))) \leq 15$~pc,} 
\State{the source $i \in$ UBC~7},
%\EndIf
\Else
\State{the source $i \in$ halo.}
\EndIf
\EndIf
%\EndIf
\end{algorithmic}
%\end{description}

The border between the cluster and halo members was selected to be equal to 15~pc, tentatively, based on the results of the analysis of the spatial distribution of Sample I. As a result, we related 294 stars to Cr~135, 243 stars to UBC~7, and 884 stars to the extended halo (see Fig.~\ref{fig:clu_halo_back}). 

\section{Estimation of the parameters of the clusters}
\label{sec:sys}

\begin{figure}
  \includegraphics[width=1.0\linewidth]{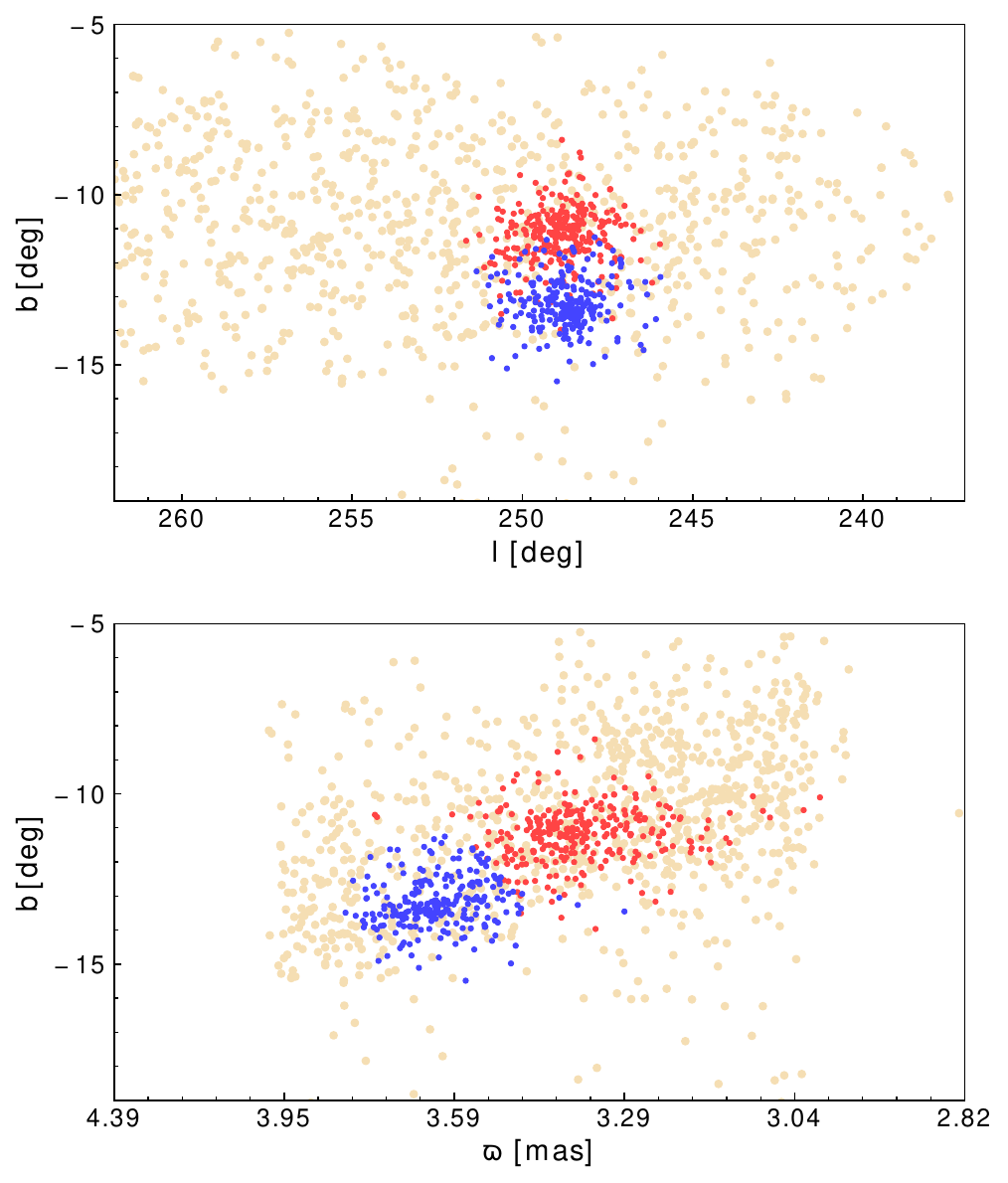}
\caption{Components of the OC system. The probable members of Cr~135 are shown with red dots, probable members of UBC~7 are indicated with blue dots, and extended halo members are shown with yellow dots. The upper panel shows the $lb$ Galactic coordinates sky plane; the  bottom panel shows the  $\varpi,\,b$ plane.}
  \label{fig:clu_halo_back}
\end{figure}

The system includes two quasi-ellipsoidal dense stellar groups that we refer to as clusters, which have similar but distinct mean space motions. They are surrounded by an extended stellar halo of irregular form and have an approximately constant stellar density in projection onto the $lb$-plane (Fig.~\ref{fig:clu_halo_back}, upper panel). This structure extends for more than 80 pc towards the Galactic anti-centre and half as much in the opposite direction. The dimensions towards and from the Galactic plane are comparable with the size of the clusters. In space, the structure is aligned along the direction joining the centres of Cr~135 and UBC~7, becoming closer to the Galactic plane the farther away it is from the Sun (see Fig.~\ref{fig:clu_halo_back}, bottom panel). 

\begin{figure}
  \includegraphics[width=1.0\linewidth]{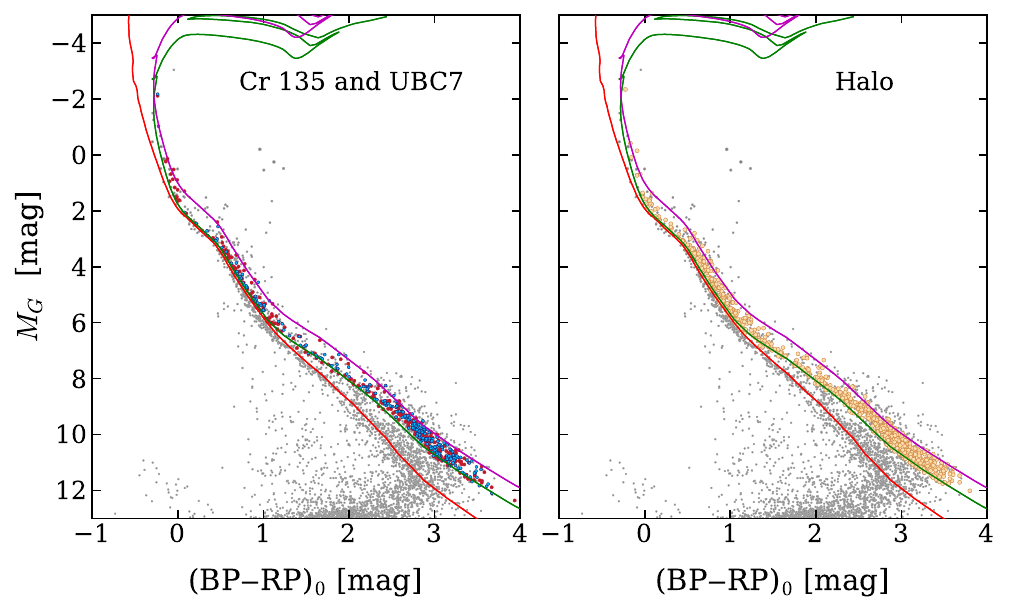}
 \caption{ Components of the OC system at the colour-magnitude diagram. Probable members of Cr~135 are shown with red dots, probable members of UBC~7 are indicated with blue dots (left panel), and probable members of the extended halo are shown with yellow dots (right panel). Grey dots are for background stars. The green line represents the isochrone for 50 Myr for single stars, and the magenta line is for the same isochrone for unresolved binary stars with identical components. The red line is for the zero-age main sequence. }
  \label{fig:zams}
\end{figure}

{The similarity of the populations of two clusters and of the population of the extended halo is illustrated by their distribution over the colour-magnitude diagram in Fig.~\ref{fig:zams}. In the figure, apparent magnitudes and reddened colours have been reduced to the absolute scale using individual parallaxes of the stars and mean reddening of the central probable members as in \hyperlink{K20}{\color{blue}{Paper~I}}. The Green line represents the Padova isochrone for 50 Myr for single stars, and the magenta line is for the same isochrone for the unresolved binary stars with identical components. The red line is for the zero-age main sequence built in \hyperlink{K20}{\color{blue}{Paper~I}} as a hot envelope of the related set of Padova isochrones of different ages. The probable members of the components of the system are represented by the same colours as in Fig.~\ref{fig:clu_halo_back}. For reference, grey dots represent each tenth background source of the same spatial region (as in Fig.~\ref{fig:widestrip}). One may notice not only the similar age of the OCs and the halo but also the similar density distribution of probable members along the isochrone, which suggests a common origin.}

The characteristics of the components of a system composed of the two OCs and their halo based on Sample I data selection are listed in  Table~\ref{tab:population}. This includes the number of probable members as defined in Section~\ref{sec:memb}, $N$, and the half-number radii of the clusters and halo, $R_{HN}$.

One may expect that the halo contains significant contamination and a random background. Its numerical estimate based on the surface density of a random background in the neighbouring regions led to $0.25 \pm 0.05$ contaminating stars per square degree. This decreased the expected actual population of the halo by 60 to 120 stars, taking into account the vagueness of its borders. 

\subsection{Counted masses}
\label{sec:countmass}

The estimate of the present-day mass of the system may be assessed based on masses of visible probable members as $\approx 900 M_\odot$ (see Table~\ref{tab:population}, $M^\Sigma$). More than half of the mass lies in the halo. There are a few probable causes of it being underestimated, such as hidden mass in low-mass stars beyond the photometric limit, members with an MP less than $60\%$, and unresolved binarity. As well, there is likely random contamination that causes present-day mass overestimation. The effect of random contamination is mostly important for the halo due to its large extent. We provide a numerical estimate for the expected contamination based on the surface density of the random background in the neighbouring regions as obtained in Sec.~\ref{sec:sys}. Filtering out the expected background reduced the mass of the halo by 40 to $70 M_\odot$. For the clusters, the halo itself should be considered a contaminating background; the estimate of the mean density for it is $3.6 \pm 0.2$ stars per square degree. This provides an upper estimate for a contaminating mass of $\approx 27 M_\odot$ for Cr~135 and $\approx 17 M_\odot$ for UBC~7.

A possible source of mass underestimation lies within the unresolved stellar multiplicity and, primarily, binarity \citep[see discussion in][]{2019ApJ...874..127B, 2021ApJ...908...60B}. The effect depends on the fraction of binary to multiple stars and their distribution over the mass fraction $q=m_B/m_A$, where $m_A$ and $m_B$ are respectively masses of the most massive and least massive components of a binary {star}.We estimated the fraction of binary stars $\alpha=N_{bin}/N_{tot}$ with mass ratio of components $q>0.4$ for the system directly from the colour-magnitude diagram  \citep[]{2019ApJ...874..127B, 2020INASR...5..351B} as $\alpha = 23\% \pm 12 \%$. The result is consistent with the binary fraction estimates carried out by \cite{Pang2023}. The estimated fraction {of binary stars} leads to the increase of the expected mass of the system by a factor of 1.1.

The complete mass of all stellar members of the system includes low-mass stars beyond the photometric completeness limit. We estimated the total hidden mass, including the unresolved components of binaries, based on the suggestion of a Kroupa-like IMF \citep{2001MNRAS.322..231K} of the system, with normalisation to the number of stars with a visible G magnitude $phot\_g\_mean\_mag\_corr<17.5$. This provided an estimate of the total mass, including unseen stars with masses down to $0.09 \msun$, that is, 1.3 to 1.4 times larger than the straightforward sum of masses of observed sources. These suggestions provide a scale of reliability of counted mass and resulted in the asymmetrical uncertainties indicated in Table~\ref{tab:population}.

In Table~\ref{tab:population}, $M^\Sigma$ is the apparent counted masses of the $60\%$-probability members. The indicated confidence limits in the table represent an extreme possible increase and decrease of this value due to taking into account hidden members or filtering out a maximum possible random contamination. The mean mass of a star is shown by $\overline{M_i}$. The mean mass of the probable members of Cr~135 $\overline{M_i}$ exceeds the mean mass of the probable members of UBC~7 and that of the halo (which are similar; see Table~\ref{tab:population}). This may be a result of mass segregation.

While the halo as a whole is more massive than both clusters together, its density is much lower, below $0.001 \msun/pc^3$. The mean stellar density within 15 pc from the centre of Cr~135 and UBC~7 is 20 to 30~times larger, and in the centre of the cores of the clusters, it is up to around $3.5 \msun/pc^3$. 

\begin{figure}
\includegraphics[width=0.99 \linewidth]{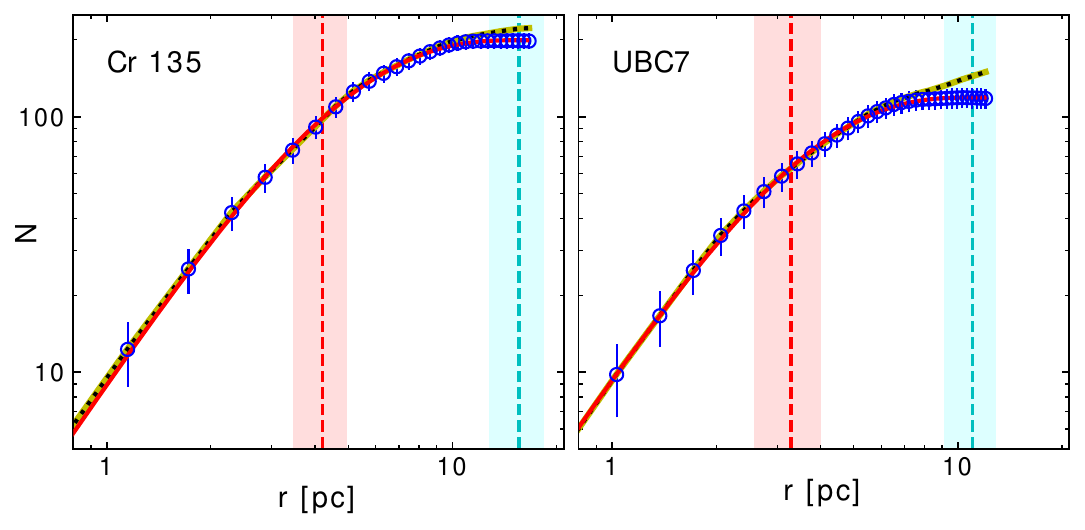}  
\caption{Fits of the King profile according to Eq. \ref{eq:kingi} for the observed radial density profiles, based on the confident members of the clusters. The curves present distributions including the residual background stars (black dots on the yellow lines), the distributions corrected for the background (circles with statistical error bars), and the fitted King profiles (solid red curves). The vertical dashed lines show the core and tidal radii ($r_{\rm c}$ and $r_{\rm t}$). The colour bands show errors in the radii.}
  \label{fig:king-fit}
\end{figure}

\subsection{Tidal parameters of substructures of the Cr~135 and UBC~7}
\label{sec:tidal}

The spatial structure of an OC can be standardised with the help of a well-known empirical model  proposed by \cite{1962AJ.....67..471K}. The model parameterises the radial surface density profile of a cluster in terms of its core and tidal (total) radii $r_c$ and $r_t$. In the integrated form, the profile can be represented with the following equation:
\begin{equation}\label{eq:kingi}
 \begin{split}
  n(r) = \pi\, r_{c}^{2}k\, 
          &\left\{\ln[1+(r/r_{c})^{2}]-4\frac{\left[1+(r/r_{c})^{2}\right]^{1/2}-1}{\left[1+(r_{t}/r_{c})^{2}\right]^{1/2}}+\right.\\
          &\left.+\frac{(r/r_{c})^{2}}{1+(r_{t}/r_{c})^{2}}\right\}\,,
 \end{split}
\end{equation}
where $n(r)$ is the number of stars projected onto the ideal (tangential) plane within a circle with radius $r$ and $k$ is a normalising coefficient. 

The observed profiles were counted based on the sources of Sample I with brightness limitation $phot\_g\_mean\_mag\_corr<18.0$. In order to avoid mutual contamination, we evaluated the residual background level produced by the neighbouring population via estimation of the average surface density produced by the sample stars in the neighbouring circular area outside the visual limit of the respective cluster. The fit and determination of the parameters was carried out using the MPFIT procedure from the \cite{2009ASPC..411..251M} IDL library.

\begin{figure}
\centering
\includegraphics[width=0.80 \linewidth]{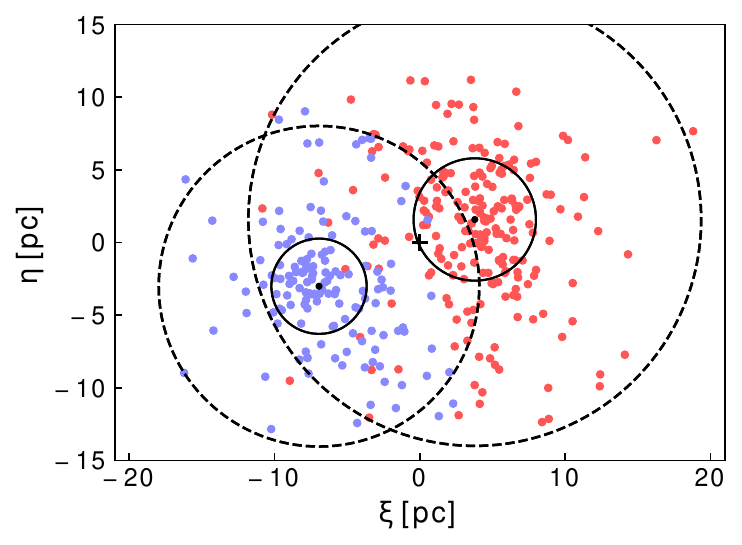}
  \caption{Open clusters Cr~135 and UBC~7 in the tangential plane contacting the celestial sphere at the common centre of the mass of the clusters. The red dots show probable members of Cr~135, and blue dots show probable members of UBC~7. The circles show the core and tidal radii. Black dots show the centres of the clusters. The ``+'' sign shows the centre of mass, which is assumed to be at 292 pc from the Sun. Linear coordinates $\xi$ and $\eta$ are parallel to equatorial ones.}
  \label{fig:tidal-map}
 \end{figure}

Figure~\ref{fig:king-fit} shows the empirical profile fits for our working samples. Figure~\ref{fig:tidal-map} illustrates how reasonably computed tidal parameters outline the general structure of our {pair of star clusters}. As one can see, the core radii clearly mark off the densest parts of the {star} clusters. In 3D, both clusters are touching at their tidal limits. We regard this as evidence of the physical relation of both clusters residing in the common potential well. In fact it is clear that the bodies that are currently recognised as separate OCs themselves are rather double cluster cores immersed in the common halo.

In Table~\ref{tab:population} we show the derived profile parameters. The fit procedure works well in our case, and the profile correction for residual outer background plays a critical role for the correct determination of the tidal radii. 

\begin{table}[htb] 
\caption{Global population characteristics and tidal parameters of the system.} 
\label{tab:population}
\begin{tabular}{l|c|c|c}
\hline
\hline
Parameter & Cr~135 & UBC~7 & Halo\\
\hline

$N$ & 294& 243 & 884\\
 $R_{HN}$, pc  & 6.4 $\pm$ 0.2 & 5.8 $\pm$ 0.2  &42.8 $\pm$ 2.1  \\
 $\overline{M_i},\,\msun$ & 0.74 $\pm$ 0.04 & 0.60 $\pm$ 0.04 & 0.63 $\pm$ 0.02 \\
 $M^\Sigma,\,\msun$ & 218$^{+87} \rm _{-27}$ & 146$^{+58} \rm _{-17}$ & 558$^{+220} \rm _{-70}$ \\
% $M^\Sigma_{corr},\,\msun$& 440 $\pm$ 90 & 324 $\pm$ 64 & 1060 $\pm$ 300 \\
$r_{\rm c}$, pc & 4.21 $\pm$ 0.75 & 3.28 $\pm$ 0.72 & --\\
$r_{\rm t}$, pc & 15.6 $\pm$ 2.3& 11.0 $\pm$ 1.9 & --\\
$m_{\rm t},\,\msun$ & 770 $\pm$ 340 & 270 $\pm$ 140 & --\\
\hline
\end{tabular}
%\vspace*{-5mm}
\\
\end{table}

In order to understand whether one can observe a footprint of cluster gravitational interaction in the observed structure of the {pair of star} clusters, we tried to compare tidal parameters for the component profiles counted in sectors opened at different directions. As we found no systematic difference is observed, the derived tidal parameters vary within the fit errors only. 

Knowledge regarding cluster structure tidal parameters can be used for independent mass estimates of an OC. To determine the tidal mass $m_{t}$, we followed \cite{1962AJ.....67..471K}, \cite{2010A&A...524A..62E} and \cite{2023A&A...672A.187J} and used a condition for the balance of gravitational forces between the Galaxy and the OC:

\begin{equation}\label{eq:jamas}
m_{t}=\frac{r_J^3}{G}(4-\beta^2)\Omega^2,
\end{equation}

where $r_J$ is the Jacobi radius (distance from the cluster centre to Lagrange points $L_1,$ or $L_2$, where its own gravity is equal to the Galaxy field), $\varOmega$ is the angular velocity of rotation of the Galaxy at the cluster's galactocentric distance $R$, $\beta$ is the ratio of epicyclic $\kappa$ and rotational $\varOmega$ frequencies, and $G$ is the gravitational constant.

The relation between Jacoby and tidal radii for realistic clusters in the case of the three-component Plummer-Kuzmin model of Galactic potential \citep{1975PASJ...27..533M} was studied using $N$-body calculations by \cite{2010A&A...524A..62E}. They showed that the ratio $r_t/r_J$ depends on the cluster position on the sky (mainly on Galactic latitude) and on their age. Its typical value at low latitudes of $|b|<20\degr$ computed by \cite{2010A&A...524A..62E} for age of cluster model of 0.62 Gyr is lower than $r_t/r_J\approx 1.19$. For the location of our clusters at $R\approx 8.29$ kpc, the rotation parameters computed from the Plummer-Kuzmin model are equal to $\beta\approx 1.45$, $\varOmega\approx 28.20$ km/s/kpc. The respective values of the tidal masses are shown in Table~\ref{tab:population}, and the formal uncertainty was estimated via propagation of the error in $r_t$.  Since our clusters are much younger than the lower age limit of the \cite{2010A&A...524A..62E} model and since we observed that the ratio tends to decrease at younger ages, we believe that the found values manifest the upper limits of the respective tidal masses.

The mass estimate obtained for UBC~7 via correction of counted mass for hidden mass ($\approx 204 M_\odot$; see Table~\ref{tab:population}) agrees rather well with the tidal mass ($270 \pm 140 M_\odot$). The tidal mass for Cr~135 ($770 \pm 340 M_\odot$) exceeds the estimate obtained from the counted mass even after correcting for the expected hidden mass ($\approx 305 M_\odot$) and taking into account the large uncertainty. This may be related to difficulties with the treatment of a populated halo. In addition, as noted previously, the found values probably manifest the upper limits of respective tidal masses. 

One may also compare present-day tidal radii with the results of the dynamical evolution modelling (Fig.~\ref{fig:tid} for the present  time $T=50$ Myr). The model predicts very similar tidal radii for the two OCs, about 12~pc for Cr~135, which is slightly less than for UBC~7. Taking into account the errors, the results match the tidal radii obtained from observations quite satisfactorily, especially for UBC~7. For Cr~135, the estimated tidal radius ($r_t=15.6 \pm 2.6$~pc; Table~\ref{tab:population}) is larger than what the dynamical evolution model predicts.

%%%%%%%%%%%%%%%%%%%%%%%%%%%%%%%%%%%%%%%%%%%%%%%%%%%%%%%%%%%%%%%%%%%%%

%%%%%%%%%%%%%%%%%%%%%%%%%%%%%%%%%%%%%%%%%%%%%%%%%%%%%%%%%%%%%%%%%%%%%
\section{Discussion} 
\label{sec:disc}

\subsection{Cumulative star number and proper motion distributions inside the modelled clusters}

\begin{figure}
    \includegraphics[width=1.0 \linewidth]{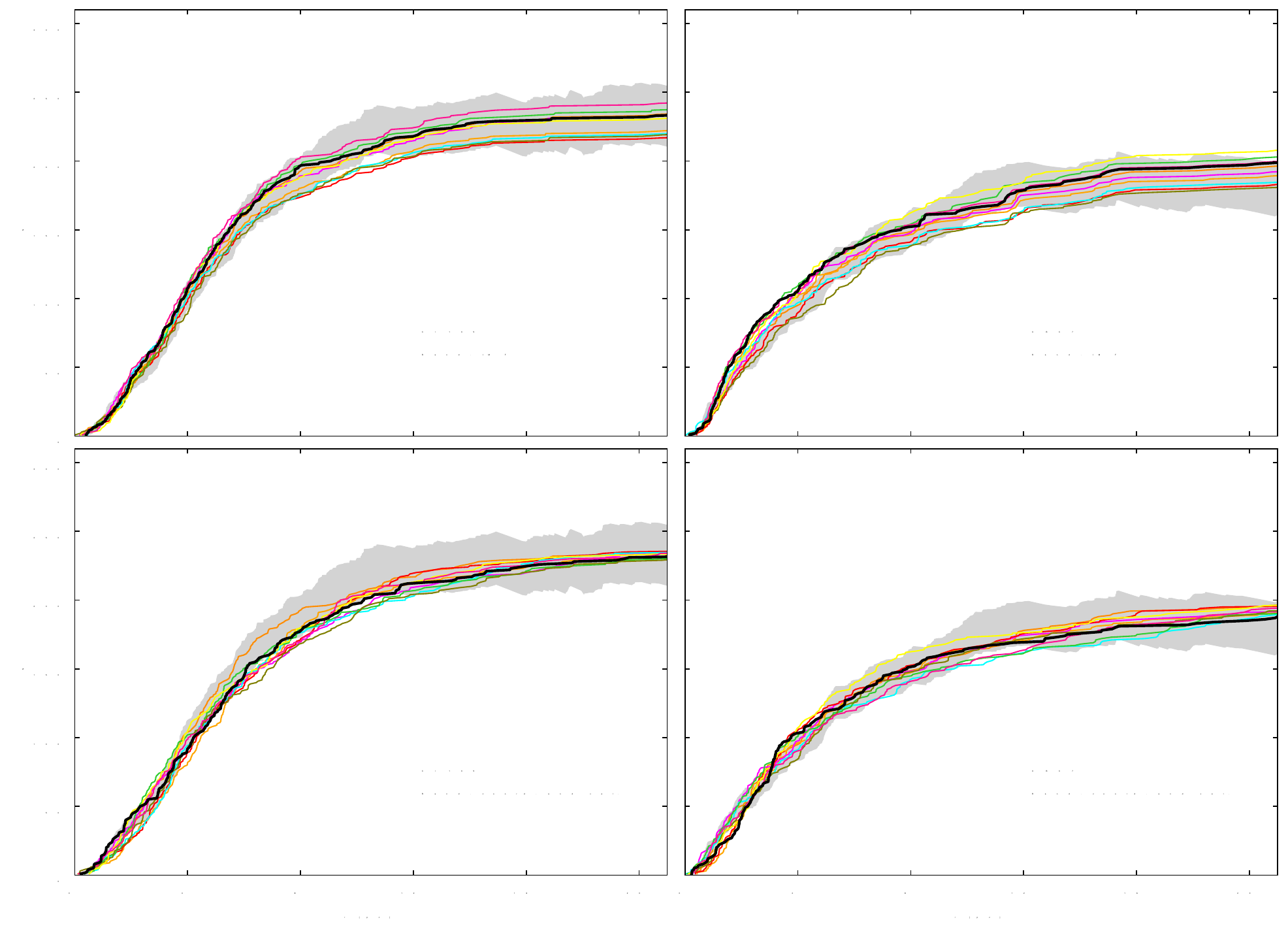}    
    \caption{Cumulative number distribution of stars compared against the projected radius for Cr~135 (left panels) and for UBC~7 (right panels). The coloured lines represent different $N$-body randomisations. The \textit{upper panel} presents the IMF sampling, and the \textit{bottom panels} show the PV sampling. The shaded grey regions show the observational data with uncertainties.}
    \label{fig:imf-col}
\end{figure}

In this section, we present a comparison of the present-day radial cumulative star count obtained from our basic model with initial conditions presented in Table \ref{tab:data1} with the current observational data (see Section \ref{sec:tidal}, Fig. \ref{fig:king-fit} with Table \ref{tab:population}). In Fig. \ref{fig:imf-col}, the total cumulative number distribution (CND) of cluster stars excluding the stellar background but including the observational errors in number count are presented as a grey zone. For Cr~135 and UBC 7, we used the fit range within $0 < r < 20$ pc. We set such a large distance limit in order to exceed the OCs' current Jacobi radii (see Table~\ref{tab:population}). We present our basic model as a thick black line in these plots, which was obtained directly from $N$-body simulations. Because of the magnitude limit and incompleteness of the faint sources in the observational data due to the \textit{Gaia} satellite specifications, we selected from the numerical models only the stars that are within the specific observed stellar mass range: from 0.28 to 4.0 $\msun$. In this case, our model (black line) has been adopted as an observed CND of stars for both clusters.

After the physical parameters of the model clusters were fixed, we carried out an additional set of 2 $\times$ 9 = 18 random numerical realisations of the same physical model. In the first nine realisations, we used different samples of the initial mass function (IMF) of the clusters, while the initial positions and velocities of the stars were fixed. In the second set of nine realisations, we randomised the initial positions and velocities (PVs) of the stars, while the IMF distribution was fixed. This enabled us to estimate the possible uncertainties directly connected with the randomisation of the initial individual stellar masses and PVs phase-space distributions. 

In Fig.~\ref{fig:imf-col}, all the coloured lines are inside the grey area, especially in the central 8 pc, which is where 80\% of the stars reside. However, our plots show that half of the coloured lines are on the edge of the grey area, around the 10-11 pc. We interpret this as an indication of a somewhat smaller concentration of the real initial OCs' PV model distribution compared to our initially selected numerical values. Taking into account the OCs' centres of mass uncertainties obtained from the 18 samples, we got the next standard deviation in the three Cartesian Galactic coordinates for Cr 135 ($\pm2.70$, $\pm2.50$, and $\pm1.86$ pc) and for UBC 7 ($\pm1.85$, $\pm1.41$, and $\pm0.96$ pc). As can be seen, we got slightly larger deviations for Cr 135 as compared to UBC~7.

\begin{figure}
    \includegraphics[width=1.0 \linewidth]{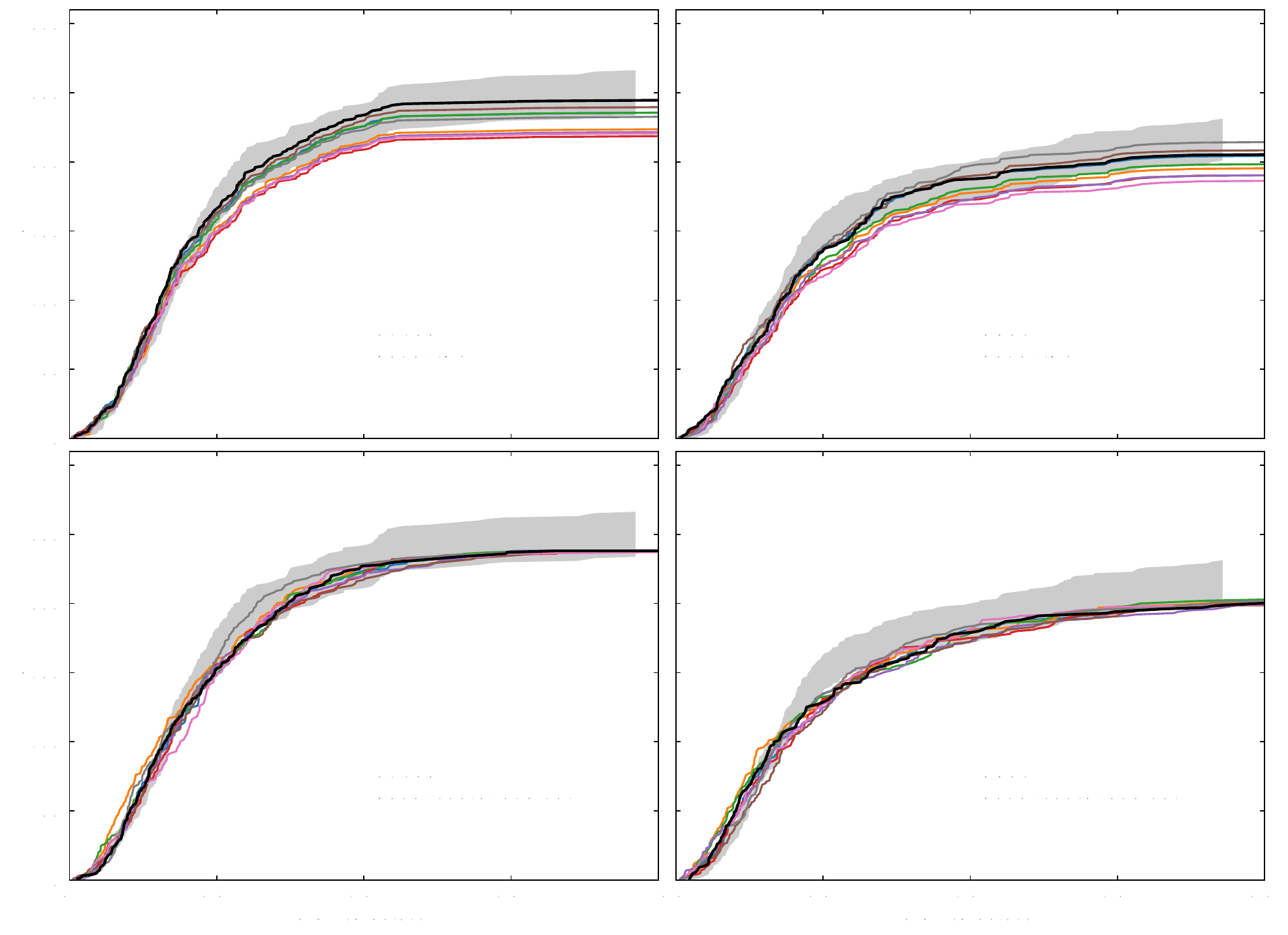}    
    \caption{Same as in Fig.\,\ref{fig:imf-col} but compared against the residual magnitude of the proper motion.\label{fig:imf-col-pm}}
\end{figure}

In Fig.~\ref{fig:imf-col-pm}, we present the same CNDs but compare them against the residual proper motion magnitude inside the clusters. As can be seen from the plots, the proper motion velocity components of the stars are also quite consistent with the observed distributions and error bars. Together with the distance CNDs, these results show that our numerical modelling is able to reproduce the currently observed 6D phase-space distributions with a high level of confidence. 

%%%%%%%%%%%%%%%%%%%%%%%%%%%%%%%%%%%%%%%%%%%%%%%%%%%%%%%%%%%%%%%%%%%%%

\subsection{Origin of the extended halo}

In Section 3, we pointed out the presence of a halo surrounding both clusters with a mean density of $3.6\pm0.2$ stars per square degree. The halo stars are not in our current $N$-body simulations because we started with two King spheres in virial equilibrium that are accurate according to mutual attraction and the Galactic tidal field. 

An explanation for the halo can be found in the cluster formation scenario with low star-formation efficiency~\citep[SFE; see][and references therein]{2021A&A...654A..53S}. It suggests that most of the gas is expelled abruptly from the star-forming region due to stellar feedback from massive OB stars. A sudden loss of approximate equilibrium in the gas-stars system leads to so-called violent relaxation followed by star ejection. 

\begin{figure}
    \includegraphics[width=1.0 \linewidth]{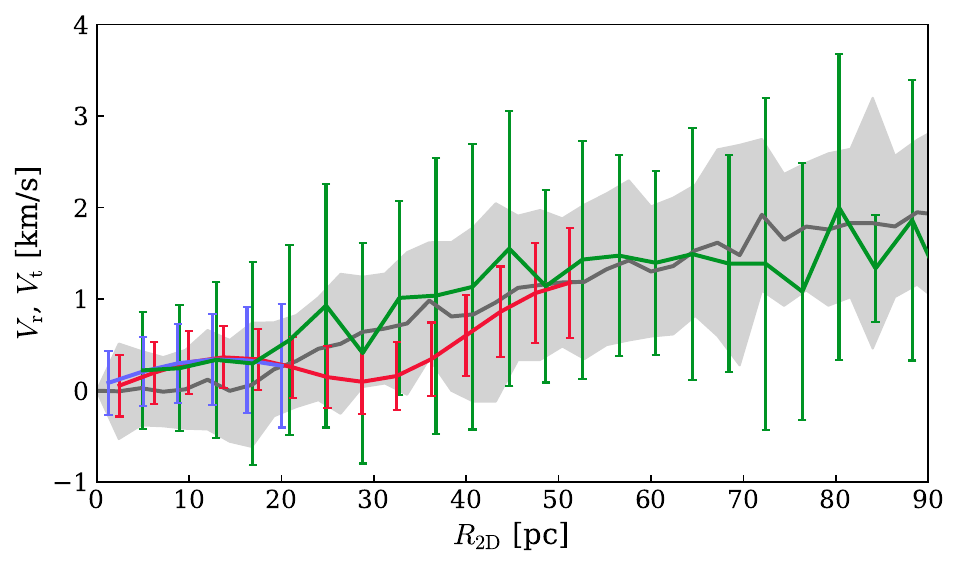}    
    \caption{Outwards motion of the corona stars with respect to the {pair of star clusters'} centre of mass. Smoothed magnitudes $V_{\rm r}$ of residual line-of-sight velocities of receding (approaching) stars from Gaia~DR3 are shown by red (blue) lines. The mean residual tangential velocities in the picture plane $V_{\rm t}$ are shown by the green line. The grey line shows the mean tangential components of the runaway velocities obtained in $N$-body simulations for the Dehnen $\gamma=0$ initial stellar profile with SFE\,$=$\,0.05 after 50\,Myr of evolution \citep{2021A&A...654A..53S}, where the shadow represents the uncertainty.
    \label{fig:halo-scattering}}
\end{figure}

To find evidence in favour of this scenario, we evaluated the net residual velocities of the stars with respect to the centre of the {pair of star clusters} in the line-of-sight and tangential directions. The obtained net line-of-sight velocities of the receding and approaching stars are shown in Fig.~\ref{fig:halo-scattering}. In the figure, the mean residual radial and tangential velocities of the cluster members (red, blue, and green lines) are compared with low-SFE cluster values (grey line), which were obtained using data by \cite{2021A&A...654A..53S}. In this paper, we used the Dehnen $\gamma=0$ initial stellar profile with SFE\,$=$\,0.05 after 50\,Myr of evolution, which is presented for the first time. This choice of the model was determined by the bound mass of the numerical cluster, which should be comparable to the mass of our {pair of star clusters}. We calculated the runaway velocities of stars from the centre of mass of the pair of clusters with the projection on the sky plane. The mean value of the velocities is shown with the grey line in Fig.~\ref{fig:halo-scattering}, and the shadow represents its standard deviation. The agreement between the net tangential observed velocity and the mean radial velocity in the simulations is excellent, supporting the scenario of low-SFE formation {of the pair of star clusters}.

%%%%%%%%%%%%%%%%%%%%%%%%%%%%%%%%%%%%%%%%%%%%%%%%%%%%%%%%%%%%%%%%%%%%%

\section{Conclusions} 
\label{sec:conc}

We performed star-by-star dynamical modelling of the evolution of the two OCs Collinder 135 and UBC 7 from their supposed initial state (mean coordinates and velocities) to their present-day state. The models were compared with the data obtained based on the performed census of membership of the two clusters with Gaia DR3 data. In particular, we used CNDs of the cluster members over the distance from their centres from observations to adjust the initial parameters of the clusters, such as the number of stars and their concentration parameter in the model. We selected these parameters in order to reproduce well the present-day state and distribution of the members of Cr~135 and UBC~7 in 6D space (coordinates, velocities). 

Our research supports the potentially shared origin of the Cr~135 and UBC~7 OCs. By integrating the dynamical co-evolution of these clusters within an external Galactic potential, we obtained results that align closely with the internal stellar density and velocity distributions of the clusters. In our parameter-fitting procedure for the clusters, we focused solely on the resulting stellar CNDs (Fig.~\ref{fig:imf-col}) at 50 Myr post cluster formation. The model's resulting profiles for the residual proper motions and the Gaia DR3 population of the clusters show a strong agreement without the need for any additional fitting, Fig.~\ref{fig:imf-col-pm}.

The star-by-star dynamical evolution model shows the future of the stellar populations of Cr 135 and UBC 7 (including the existence of the stellar remnants, such as C/O and O/N white dwarfs). Due to the limited number of stars (only a few hundred) in both clusters, their 50 Myr-evolved stellar populations have a lack of neutron stars and black holes. 

Based on our comprehensive dynamical modelling with the up-to-date stellar evolution prescription, we are able to predict the future $\sim$100 Myr evolution of both clusters with a good degree confidence. In our models, the UBC 7 cluster lives significantly longer compared to the Cr 135 cluster, Fig. \ref{fig:tid}. This is not surprising given the fact that, according to our numerical modelling, UBC 7 is denser (see Table~\ref{tab:data1}) than Cr 135. 

During our investigation, we found strong evidence of the presence of the common stellar halo in the Cr~135 and UBC~7 OC system, and we estimated its characteristics. The members of the halo demonstrate net residual velocities of stars with respect to the centre of the {pair of star clusters} that are compliant with the mean radial velocity of particles of $N$-body simulations of a low-SFE cluster stimulated by violent relaxation after gas loss. 

Detailed common dynamical modelling of the OCs and the extended halo system was beyond of the scope of our current paper. Such modelling, however, will definitely be of interest and valuable for future investigations. The presence of such an extended halo can also have an impact on the clusters' cumulative mass distributions and internal proper motions. 

%%%%%%%%%%%%%%%%%%%%%%%%%%%%%%%%%%%%%%%%%%%%%%%%%%%%%%%%%%%%%%%%%%%%%
\begin{acknowledgements}

The authors thank the anonymous referee for a very constructive report and suggestions that helped significantly improve the quality of the manuscript. 

The work of MI, PB and OS was partially supported under the special program ``Long-term program of support of the Ukrainian research teams at the PAS Polish Academy of Sciences carried out in collaboration with the US National Academy of Sciences with the financial support of external partner'', PAN.BFB.S.BWZ.329.022.2023. 

The work of PB was also supported by the Volkswagen Foundation under the special stipend No.~9D154. 

The work of CO was funded by the Science Committee of the Ministry of Science and Higher Education of the Republic of Kazakhstan, Grant No. BR20280974.

This research is funded by the Science Committee of the Ministry of Science and Higher Education of the Republic of Kazakhstan, Grant No. BR21881880.

This work has made use of data from the European Space Agency (ESA) mission {\it Gaia} (\url{https://www.cosmos.esa.int/gaia}), processed by the {\it Gaia} Data Processing and Analysis Consortium (DPAC, \url{https://www.cosmos.esa.int/web/gaia/dpac/consortium}). Funding for the DPAC has been provided by national institutions, in particular the institutions participating in the {\it Gaia} Multilateral Agreement. The use of TOPCAT, an interactive graphical viewer and editor for tabular data \citep{2005ASPC..347...29T}, is acknowledged.

\end{acknowledgements}

%%%%%%%%%%%%%%%%%%%%%%%%%%%%%%%%%%%%%%%%%%%%%%%%%%%%%%%%%%%%%%%%%%%%%

%%%%%%%%%%%%%%%%%%%%%%%%%%%%%%%%%%%%%%%%%%%%%%%%%%%%%%%%%%%%%%%%%%%%%

\normalfont

\bibliographystyle{aa}  % style aa.bst
\bibliography{main}   % your references Yourfile.bib

\begin{appendix}

\section{Calculation of probabilities of membership to the group of Cr~135 and UBC~7}
\label{memb}

We adjusted our approach to selecting probable members of a system based on data obtained in our previous investigation. That is, in \hyperlink{K20}{\color{blue}{Paper~I}}, we found the number of stars in a close vicinity to the considered pair of OCs with space parameters suitable to one cluster and kinematic parameters of another cluster. These stars were not included in the previous strict selection. However, we interpreted these stars as being probable members of a system and further computed their joint probability based on parallax and their joint kinematic probability by taking into account the parameters of both clusters.   

\subsection{Kinematic probability}

As in \hyperlink{K20}{\color{blue}{Paper~I}}, we found a halo of stars around the Cr~135 and UBC~7 pair with a location and kinematics combination that cannot be attributed to only one of the two clusters but rather to their system. This is why for the selection of space of coordinates and velocities, we did not distinguish between the two clusters but rather chose stars with both velocities and parallaxes sufficiently close to those of either of the two clusters: Cr~135 or UBC~7. We describe the details of the selection process below.

Provided that our intention was to search for coeval stars probably related to the pair of clusters in an extended region of sky, the kinematical selection principles needed to be respectively changed from 2D to 3D velocities to account for the projection effect. We used the modification of the convergent point method described by \cite{2011A&A...531A..92R}. Of the initial dataset including 116975 stars, radial velocities were available only for 4575 sources. Thus, we selected probable members related to the system based on the components of their tangential velocity. 

We expected that the members related to Cr~135 and UBC~7 follow certain distributions with respect to the mean 6D phase-space parameters cited in Table~\ref{tab:data_dr2_3} (based on Gaia DR3). The respective space velocities were obtained as 
\begin{equation}
    \begin{bmatrix} 
    U^i\\
    V^i\\
    W^i
    \end{bmatrix} = V_r^i \cdot \begin{bmatrix}
    \cos  l^i \cos b^i\\
    \sin l^i \cos b^i\\
    \sin b^i
    \end{bmatrix} +\frac{\kappa \mu_l^i}{\varpi^i} \cdot
    \begin{bmatrix}
    -\sin l^i \\
    \cos l^i \\
    0
    \end{bmatrix} + \frac{\kappa \mu_b^i}{\varpi^i} \cdot
    \begin{bmatrix}
    -\cos l^i \sin b^i \\
    -\sin l^i \sin b^i \\
    \cos b^i \\
    \end{bmatrix},
\end{equation}

\noindent where $\kappa=4.74047$ is the transformation factor from 1 mas/yr at 1 kpc to 1 km/s; the indices are $i=1$ for Cr~135 and $i=2$ for UBC~7; the mean values for coordinates, parallaxes, proper motions, and radial velocities are from Table~\ref{tab:data_dr2_3}. %The obtained values for heliocentric space velocities are listed in Table~\ref{tab:datavel}.

The kinematic probability $P^k_\textrm{kin}$ for a $k^\textrm{th}$ star to belong to the investigated system is defined as

\begin{equation}
%\begin{split}
P^k_\textrm{kin}=\max 
\begin{cases}
 \exp\left\{-\frac{1}{4}\left[\left(\frac{\mu^k_{l}-\mu_l^{exp_1^k}}{\varepsilon_{\mu_l}}\right)^2+\left(\frac{\mu^k_{b}-\mu_b^{exp_1^k}}{\varepsilon_{\mu_b}}\right)^2  \right] \right\}, \\ \exp\left\{-\frac{1}{4}\left[\left(\frac{\mu^k_{l}-\mu_l^{exp_2^k}}{\varepsilon_{\mu_l}}\right)^2+\left(\frac{\mu^k_{b}-\mu_b^{exp_2^k}}{\varepsilon_{\mu_b}}\right)^2  \right] \right\} 

\end{cases}
%\end{split}
\end{equation}

\noindent where 
%\begin{equation}
\begin{align}
%\begin{split}
\mu_l^{exp_i^k}=&(-\sin l_k \cdot U_i + \cos l_k \cdot V_i)/(\kappa/\varpi_k); \nonumber\\
\mu_b^{exp_i^k}=& (-\cos l_k \sin b_k \cdot U_i - \sin l_k \sin b_k \cdot V_i + \cos b_k \cdot W_i)/\\
& (\kappa/\varpi_k) \nonumber    
%\end{split}
%\end{equation}
\end{align}

\noindent are values of proper motion in $(l,b)$ for given coordinates $l_k, b_k$, and $\varpi_k$ expected for the stars sharing space motion $(U,V,W)_1, (U,V,W)_2$ of Cr~135 and UBC~7, respectively.
Scattering parameters of the proper motion $\varepsilon_{\mu_l}, \varepsilon_{\mu_b}$ were set to 1.0~mas/yr (about 2 km/s in tangential velocity). 
%This restriction is motivated by intention to avoid contamination from nearby stellar groups.

\subsection{Parallax probability}

In the tangential plane, we limited our search of members by the coordinates selected in Sec.~\ref{sec:memb}, but we did not discriminate between the stars by their position within this scope. The approach for selection by parallaxes is as follows: We used the probability distribution based on the mean parallaxes of Cr~135 and UBC~7 from Table~\ref{tab:data_dr2_3} and the notion of mean dispersion of parallaxes of their core members. We intentionally restricted this value to avoid contamination with peripheral members of neighbouring clusters (primarily NGC~2451B).   
The resulting formula for the probability of membership based on parallax value and parallax error is as follows:

\begin{equation}
P^k_\varpi=
\begin{cases}
1, 3.23<\varpi_k<3.66 \\
\exp\left\{-\frac{1}{2}\left(\frac{\varpi^k-3.23}{\varepsilon_\varpi^k}\right)^2   \right\}, \varpi_k \leq 3.23 \\
\exp\left\{-\frac{1}{2}\left(\frac{\varpi^k-3.66}{\varepsilon_\varpi^k}\right)^2   \right\}, \varpi_k \geq 3.66\,.
\end{cases}
\end{equation}

The selected scheme assigns $P^k_\varpi=1$ to all stars in the line-of-sight between the centres of the pair of clusters and within about 7.5~pc (in naive approach $d^k[pc]=1000/\varpi_k$) to and from each centre. The wings of the probability distribution were defined by 
\begin{equation*}
\varepsilon_\varpi^k=\sqrt{\sigma_\varpi^2+\sigma_{\varpi_k}^2}\,,   
\end{equation*}

\noindent where $\sigma_\varpi^2=0.25$~mas is the approximate mean dispersion of parallaxes of the joint ensemble of reliable members of Cr~135 and UBC~7, and $\sigma_{\varpi_k}^2$ is the nominal Gaia EDR3 parallax error for each source.

\subsection{Photometric probability}

Isochrones for 50 Myr (according to \hyperlink{K20}{\color{blue}{Paper~I}}) for \textit{Gaia} DR3 passbands from \cite{2018A&A...619A.180M} were obtained from the Padova webserver CMD3.7\,\footnote{\url{http://stev.oapd.inaf.it/cmd}} based on the calculations by \cite{2012MNRAS.427..127B} for solar metallicity $Z=0.0152$. A recent investigation by \cite{2021MNRAS.503.3279S} has confirmed close to solar abundance for Cr~135 ($[M/H]=-0.011$).
We took into account the relation between $A_{G}/E({\rm BP-RP})$ and $(BP-RP)_0$ of a star and $A_0$ (interstellar extinction at $\lambda=550$~nm) from \cite{2018A&A...616A..10G} to adjust isochrones for photometric probability calculation at a fixed value of $A_0=0.1^m$, \hyperlink{K20}{\color{blue}{Paper~I}}.   

The photometric probability $P^k_\textrm{ph}$ was computed in the CMD ($G, BP-RP$) as a maximum of probabilities and was calculated separately for Cr~135 (cluster 1) and UBC~7 (cluster 2) as follows. The photometric probability was set as equal to one for the stars occupying the CMD in an area between the isochrones for single stars and for unresolved binaries of equal mass. For all the other stars, the photometric probability depended on a difference between magnitudes:

\begin{equation}
\Delta G^i=\min \left(|G^i-G^b|, |G^i-G^f|\right),
\end{equation}
where $G^b, G^f$ are respectively brighter and fainter values of magnitude limiting the domain between the isochrones for single stars and for unresolved binaries of equal mass for a given colour $BP-RP$: 
\begin{equation}
\begin{split}
    P^k_\textrm{ph}=\max \left( \exp\left\{-\frac{1}{2}\left(\frac{\Delta G_1^k}{\varepsilon^k_{G_1}}\right)^2   \right\},  \exp\left\{-\frac{1}{2}\left(\frac{\Delta G_2^k}{\varepsilon^k_{G_2}}\right)^2   \right\} \right).
\end{split}
\end{equation} 
The value of $\varepsilon^i_{\rm ph}$ was defined from the mean dispersion of parallaxes of the joint ensemble of reliable members of Cr~135 and UBC~7, which is equal to $0.25$~mas, and the individual photometric error $\sigma^i_G$ estimated by the flux and flux error in G.

\end{appendix}

%%%%%%%%%%%%%%%%%%%%%%%%%%%%%%%%%%%%%%%%%%%%%%%%%%%%%%%%%%%%%%%%%%%%%

%%%%%%%%%%%%%%%%%%%%%%%%%%%%%%%%%%%%%%%%%%%%%%%%%%%%%%%%%%%%%%%%%%%%%
\end{document}